\documentclass[a4paper,12pt]{article}
\usepackage {latexsym}
\usepackage {amssymb}
\usepackage {amsfonts}
\usepackage {indentfirst}
\fussy                             

\begin{document}

\title{\bf Effect of matter motion and polarization in neutrino flavour
oscillations}

\author{\bf A. Grigoriev,\and \bf A. Lobanov,\and
\bf A. Studenikin\thanks{E-mail: studenik@srd.sinp.msu.ru}}
\date{}
\maketitle
\begin{center}
{\em Moscow State University,\\
Department of Theoretical physics, \\
$119992$  Moscow, Russia}
\end{center}
\begin{abstract}
{The Lorentz invariant formalism for description of neutrino
flavour oscillations in moving and polarized matter is developed.
It is shown that the neutrino effective potential, which
determines the effective mass difference between neutrinos in
matter can be significantly changed by relativistic motion of
matter. In the case of matter motion parallel to neutrino
propagation, matter effects in neutrino flavour oscillations are
suppressed. In the case of relativistic motion of matter in the
opposite direction substantial increase of effects of matter in
neutrino oscillations is predicted. The dependence of the matter
term in neutrino effective potential on the values and
correlations of the three vectors, the neutrino and matter speeds
and matter polarization, is discussed in detail.}
\end{abstract}

\vskip15pt

The problem of neutrino propagation in matter, starting from the
paper of Wolfenstein \cite {Wol78}, is still under intensive
study. One of the most famous results in this field was obtained
by Mikheyev and Smirnov \cite {MikSmi85} who showed that there is
a mechanism for resonance amplification of neutrino flavour
oscillations in matter (the MSW effect). Now it is commonly
believed that neutrino oscillations in matter can provide not only
solutions for the solar and atmospheric neutrino puzzles but also
have important applications in different astrophysical and cosmology
environments (see, for example, \cite{Raf96}-\cite{DHPPRS02} and references therein).

In this paper we should like to discuss a new phenomenon in neutrino
flavour oscillations in matter that have never been considered before neither in
textbooks and monographs on neutrino physics nor in papers in research journals.
We show below (see also \cite {GLSt0202276}) that the matter Wolfenstein term in
neutrino effective potential
can be substantially increased or suppressed in the case of matter moving with
relativistic speed. This effects can lead 
to interesting consequences for environments with  neutrino propagating
through relativistic jets of matter. 
As an example, we point out that astrophysical systems, such as quasars 
and microquasars,  are believed to emit collimated relativistic jets of 
plasma. Another important application of the studied here effect 
can be found within several models of gamma-ray bursts 
( see for a resent review \cite{Mes02}) where 
highly relativistic matter with Lorentz factor $\gamma \sim 
10^2-10^3$ is supposed to be accelerated in the fireball shocks. 

In the frame of the Lorentz invariant approach
(see \cite {GLSt0202276},\cite{ELSt99}-\cite{DELSt01}) to description of the
neutrino flavour and spin oscillations in matter
we get below the effective neutrino potential which contributes to the
Hamiltonian describing neutrino evolution in the case of matter motion with
arbitrary (also relativistic) total speed.  We show here that the matter
motion could lead to sufficient change in the neutrino flavour oscillation
probabilities and other neutrino oscillation parameters
(like the effective mixing angle and oscillation length) if matter is
moving with relativistic speed.  In particular, in the case of matter
moving parallel to the neutrino propagation, the Wolfenstein matter
term in the neutrino effective potential is suppressed.  Contrary
to this suppression effect, in the case of relativistic motion of
matter in the opposite direction in respect to neutrino propagation,
substational increase of effects of matter in neutrino flavour
oscillations is predicted. We also argue that effects of matter motion
have to be accounted for in the resonance conditions for the neutrino
flavour oscillations. Effects of matter polarization are also taken
into account, and the dependence of the matter term in neutrino
effective potential on the values and correlations of the three
vectors, the neutrino and matter speeds and matter polarization, are
discussed in detail.

It should be also noted here that effects of matter polarization in
neutrino oscillations were considered previously in several papers
( see, for example, \cite {NSSV97,BGN99} and references therein).
However, the procedure used in refs. \cite {NSSV97,BGN99} of
accounting for the matter polarization effect does not enable one to study
the case of matter motion with total relativistic speed.
Within our approach we can reproduce corresponding results of
\cite {NSSV97,BGN99} in the case of matter which is slowly moving or
is at rest.

Our goal is to investigate neutrino oscillations characteristics in
the case of relativistic motion of matter which is composed of different
charged and neutral background fermions, $f=e,n,p,\mu,\tau$, $\nu_e, \nu_\mu,
\nu_\tau$, etc, accounting for possible effects of matter motion and polarization.
In the general case each of the matter components is characterised by
the number density $n_{f}$, the speed,
$\vec v_f$, of the reference frame in which the mean momentum of the fermion
$f$ is zero, and fermion polarization $\lambda^{\mu}_f$. The fermion $f$
current is determined as
\begin{equation}
j^{\mu}_{f}=(n_f, n_f \vec v_f).
\label{curr}
\end{equation}
If a component of the matter $f$ is slowly moving or is at rest in the laboratory frame, $v_f \approx 0$,
the fermion current equals
\begin{equation}
j^{\mu}_{f}=(n_f, 0,0,0).
\label{curr_0}
\end{equation}
The polarization of each of the matter components $f$ is given by
\begin{equation}
\lambda^\mu_f = \left( n_f \vec \zeta_f \vec v1_f ,
n_f\vec \zeta_f \sqrt{1-v^2_f}+{n_f\vec v_f \left( \vec \zeta_f \vec v_f \right)
\over 1+\sqrt{1-v^2_f}} \right).
\end{equation}
Here vectors $\vec \zeta_f$ $( 0 \leqslant | \vec \zeta_f
|^2\leqslant 1 )$ are the mean values of polarization vectors of
background fermions $f$ in the reference frame in which the mean
momentum of fermions $f$ is zero. These vectors are determined by the
two-step averaging procedure described in \hbox{ref. \cite
{LobStuph01}.} In the case of non-moving matter component $f$,
$\vec v_f=0$, the fermion $f$ polarization is
\begin{equation}
\lambda^\mu_f = (0,n_f\vec \zeta_f).
\end{equation}

Let us suppose that at least one of the matter components $f$ is moving as a whole with relativistic speed,
$v_f \approx 1$. For simplicity, let us consider
neutrino two-flavour oscillations, e.g. $\nu_{e} \leftrightarrow \nu_{\mu}$, in matter composed of
only one component, electrons ($f=e$), moving with relativistic total speed. Generalisation for
the case of other types of neutrino conversions and different compositions and types of motion of
matter is straightforward.

The matter effect in neutrino oscillations occurs as a result of
elastic forward scattering of neutrinos on the background
fermions. In our case the difference $\Delta V$ between the
potentials $V_{e}$ and $V_{\mu}$ for  the two-flavour neutrinos is
produced by the charged current interaction of  the electron
neutrino with the background electrons. Note that the neutral
current interaction is affective in oscillations between the
active and sterile neutrinos. Concluding remarks of the paper are
devoted to this type of neutrino oscillations. We do also account
for effects of matter polarization which vanishes if there is no
preferred spin orientation of electrons. Note that these effects were
discussed in detail for neutrino flavour oscillations in
non-moving matter \cite {NSSV97} and also for neutrino spin
oscillations in moving matter  \cite {LobStuph01}. We neglect
below the momentum dependence of the charged vector boson
propagator. Then the corresponding part of the neutrino effective
Lagrangian can be written in the following form \hbox{(see \cite
{LobStuph01} and also \cite {Pal92})}
\begin{equation}
{\cal L}_{eff} =-f^{\mu}\left( {\bar {\nu}}{\gamma}_{\mu}{ 1+{\gamma}^{5}
                                  \over 2}{\nu} \right),
\end{equation}
where
 \begin{equation}
f^{\mu}=\sqrt2G_F(j^{\mu}_e-{\lambda}^{\mu}_e).
\end{equation}
This additional term in the Lagrangian modifies the Dirac equation
for neutrino:
\begin{equation} ({\gamma}_0E-{\vec {\gamma}\vec
p}-m)\psi=(\gamma_{\mu} f^{\mu})\psi .
\label{Dir}
\end{equation}
Rearranging the terms we get the neutrino dispersion relation in matter in
the following form,
\begin{equation}
E=\sqrt{{(\vec p-\vec f )}^2+m^2}+f_0.
\label{disp}
\end{equation}
The further simplification is found to occur in the limit of weak
potential $|\vec f| \ll p_0=\sqrt{{\vec p}^{\;2}+m^2}$. Thus we
get for the effective energy of the electron neutrino in the
moving and polarized matter
\begin{equation}
\matrix {\displaystyle E=\sqrt{{\vec p}^{\;2}+m^2}+U \left\{
{(1-\vec {\zeta}_e {\vec v}_e)(1-\vec {\beta}{\vec v}_e)+ }
\right. & \cr \left. {\displaystyle \sqrt{1-v^{2}_e} \left[ {\vec
{\zeta}_e {\vec \beta} -{(\vec {\beta} {\vec v_e})(\vec {\zeta}_e
{\vec v_e}) \over 1+\sqrt{1-v^2_e}}} \right] } \right\} +
O({\gamma}^{-1}_{\nu}), & \cr }
 \label{energy}
\end{equation}
where $\vec \beta$ is the neutrino speed, $\gamma_{\nu}=(1-{\beta}^2)^{-1/2}$,
and in the considered case of the two-flavour neutrino oscillations
$\nu_e \leftrightarrow \nu_\mu$ and one-component matter
 $U=\sqrt2 G_Fn_e$.

The effective electron neutrino energy in addition to the vacuum energy $p_0$ contains
the matter term which is proportional to the electron number density $n_e$.
Let us underline that $n_e$ is related to the invariant electron number
density $n_0$ given in the reference frame for which the total speed of electrons
is zero,
\begin{equation}
n_e=n_0/\sqrt{1-{v_e}^2}.
\end{equation}
There is no matter term in the muon neutrino effective energy in
the case of matter composed of only electrons. If matter is slowly
moving or is at rest, $v_e \approx 0$, eq. (\ref{energy}) includes the well
known result for the Wolfenstein matter term  \cite {Wol78}.

The important new phenomenon, as it can be seen from eq.
(\ref{energy}), is the matter term dependence on the value of the total speed
$\vec v_e$, polarization $\vec \zeta_e$ of background electrons,
neutrino speed $\vec \beta$ and correlations between these three vectors.

Now following the usual procedure for the two-flavour neutrinos, $\nu_e$ and $\nu_\mu$, with mixing
in the high-energy limit it is
possible to get the constant-density solution for the neutrino oscillation problem. Thus, the probability
of neutrino conversion $\nu_e \rightarrow \nu_\mu$ can be written in the form
\begin{equation}
P_{\nu_e \rightarrow \nu_\mu}(x)=\sin^{2} 2\theta_{eff} \sin^{2} {\pi x \over L_{eff}},
\label{ver}
\end{equation}
where $x$ is the distance travelled by neutrino in moving and polarized matter. 
The effective mixing angle, $\theta_{eff}$, and the effective oscillation length, $ L_{eff}$,
are given by
\begin{equation}
\sin^{2} 2\theta_{eff}={\Delta^{2}\sin^{2} 2\theta \over
{\Big(\Delta \cos2\theta - A\Big)^2+ \Delta^{2}\sin^{2} 2\theta}},
\label{th}
\end{equation}
\begin{equation}
L_{eff}= {2\pi \over
{\sqrt {\Big(\Delta \cos2\theta - A\Big)^2+ \Delta^{2}\sin^{2} 2\theta}}}.
\label{l}
\end{equation}
Here $\Delta = {\delta m^{2}_\nu / {2 |\vec p\,|}}$, $\,\delta
m^{2}_\nu=m^{2}_{2}-m^{2}_{1}$ is the difference of the neutrino
masses  squired, $\vec p$ is the neutrino momentum, $\theta$ is
the vacuum mixing angle and
\begin{equation}
\begin{array}{c}
A=\displaystyle\sqrt2 G_F{n_0 \over \sqrt{1-v^2_e}}\left\{(1-{\vec
\beta}{\vec v_e}) (1-\vec {\zeta}_e {\vec v_e})\right. + \\
\left.\displaystyle\sqrt{1-v^2_e} \left[ \vec {\zeta}_e {\vec
\beta} -{(\vec {\beta} {\vec v_e})(\vec {\zeta}_e {\vec v_e})
\over 1+\sqrt{1-v^2_e}} \right] \right\} . \label{A}
\end{array}
\end{equation}

One can see that the neutrino oscillation probability, $ P_{\nu_e
\rightarrow \nu_\mu}(x)$, the  mixing angle, $\theta_{eff}$, and
the oscillation length, $ L_{eff} $, exhibit dependence on the
total speed of electrons $\vec v_e$, correlation between $\vec
\beta$, $\vec v_e$ and polarization of matter $\vec \zeta_e$. The
resonance condition
\begin{equation}
 {\delta m^{2}_\nu \over {2 |\vec p\,|}}\cos 2\theta= A,
\label{res}
\end{equation}
at which the probability has unit amplitude no matter how small
the mixing angle $\theta$ is, also depends on the motion and
polarization of matter and neutrino speed. We show below that the
relativistic motion of matter could provide appearance of  the
resonance in the neutrino oscillations in certain cases when for
the given neutrino characteristics, $\delta m^{2}_\nu$, $|\vec
p\,|$ and $\theta$, and the invariant matter density at rest,
$n_0$, the resonance is impossible.

Let us consider first unpolarized matter, $\zeta_e \approx 0$. Then
the expression in the right hand side of eq. (\ref{res}) is equal to the difference
of the vector parts of neutrino $\nu_e$ and $\nu_\mu$ effective potentials
(see also \cite{GLSt0202276}),
\begin{equation}
V^V=\sqrt2G_F{n_0 \over {\sqrt {1-v^2_e}}}(1-\vec \beta \vec v_e).
\label{V^Vz0}
\end{equation}
The resonance condition (\ref{res}) can be rewritten as
\begin{equation}
 {\delta m^{2}_\nu \over {2 |\vec p|}}\cos 2\theta=
\sqrt2 G_{F} {n_{0} \over {\sqrt {1-v^{2}_e}}}(1-\vec \beta \vec
v_e). \label{res0}
\end{equation}
Note that the value of $n_{0}$ gives the matter density in the
reference frame for which the total speed of matter is zero. It
follows that the effect of matter motion in the resonance condition
depends on the factor
\begin{equation} a={ {1-\vec \beta \vec v_e}
\over {\sqrt {1-v^{2}_e}}}\,. \label{fac}
\end{equation}
The appearance
of the same factor in the matter term for neutrino spin oscillations
was shown in \cite{LobStuph01}.

If one estimates this factor for the ultra-relativistic neutrinos,
$\beta \approx 1$, and matter moving along the direction of neutrino
propagation with total speed $v_e$, then
\begin{equation}
{{1-\vec \beta \vec v_{e}} \over {\sqrt {1-v^{2}_e}}} =
\sqrt{{1-v_e} \over {1+v_e}}\,. \label{fac1}
\end{equation}
Here we suppose that the condition $\gamma_e \ll \gamma_{\nu}$ is valid,
$\gamma_e=(1-v^2_e)^{-1/2}$. For example, this condition in the case of the
neutrino mass and energy to be equal to $m_{\nu} \sim 2eV$ and
$E_{\nu}\sim10MeV$ sets a limit on the total energy of the background electrons
on the level of $E_e<2.5TeV$.
In the opposite case, when matter is moving against the direction to
the neutrino propagation, we get
\begin{equation}
{ {1-\vec \beta \vec v_e} \over {\sqrt {1-v^{2}_e}}} = \sqrt
{{1+v_e} \over {1-v_e}}\,. \label{fac2}
\end{equation}

From these estimations it follows (see \cite{GLSt0202276}) that:
1) the relativistic motion of matter, $v_e \sim 1$ , along the neutrino propagation could
provide the resonance increase of the oscillation probability if  the matter density $n_0$ is too high for the resonance
appearance in non-moving matter,
2) the relativistic motion of matter in opposite direction to  the neutrino propagation could
provide the resonance increase of the oscillation probability if  the matter density $n_0$ is too low for the resonance
appearance in non-moving matter.

Now let us discuss the case when the matter polarization effects are also
important, $\vec \zeta_e \not=0 $. Consider the axial part of the difference
of neutrino effective potentials. From the general expression, eq.(\ref{A}),
for $A=V^V+V^A$ we find that the axial part is given by
\begin{equation}
V^A=\sqrt2G_Fn_0\left\{ \vec {\zeta}_e {\vec \beta}-{(1-{\vec \beta}{\vec v_e})
(\vec {\zeta}_e {\vec v_e}) \over \sqrt{1-v^2_e}} -
{(\vec {\beta} {\vec v_e})(\vec {\zeta}_e {\vec v_e}) \over 1+\sqrt{1-v^2_e}}
\right\} .
\label{V^A}
\end{equation}
In the case of non-moving matter, $v_e \sim 0$, we find
\begin{equation}
V^A=\sqrt2G_Fn_0\vec \zeta_e \vec \beta.
\label{V^Av1}
\end{equation}
This result coincides with the corresponding expression of ref. {\cite{NSSV97}}
if one follows approximation $\vec \beta =\vec k_\nu$, where $\vec k_\nu={\vec p_\nu
/|\vec p_\nu|}$ is the unit vector in the direction of the neutrino
momentum.

We get the same result, eq. (\ref{V^Av1}), if electrons are polarized in the transverse
plane, $\vec \zeta_e \vec v_e=0$. Let us underline that due to the condition
$\vec \zeta_e \perp \vec v_e$ the axial potential $V^A$ does not
depend on the absolute value of matter speed $v_e$. In this case for the
relativistic neutrino $V^A$ is zero only if neutrino is also
propagating in the perpendicular direction to the matter polarization.
Note that this statement is valid also for the case of the
relativistic motion of matter (see below).

In an analogy with what have been
found for the matter term in the neutrino spin oscillations {\cite{LobStuph01}},
the potential $V^A$ exhibits critical dependence on the correlation between the
direction of the neutrino propagation,
$\vec k_\nu$, and the direction of the matter motion given by the unit vector
($\vec k_e=\vec v_e /v_e$).
For these two particular cases, $\vec k_\nu \vec k_e=\pm 1$, from eq. (\ref{V^A})
we get
\begin{equation}
V^A \left|_{\atop (\vec k_\nu \vec k_e)=1} \right. = \sqrt2
G_Fn_0\vec \zeta_e \vec k_\nu\sqrt{{1-v_e} \over {1+v_e}}\,,
\label{V^Av=1}
\end{equation}
and
\begin{equation}
V^A \left|_{\atop (\vec k_\nu \vec k_e)=-1} \right. = \sqrt2
G_Fn_0\vec \zeta_e \vec k_\nu\sqrt{{1+v_e} \over {1-v_e}}\,.
\label{V^Av=-1}
\end{equation}
For the neutrino and matter motion along one line
with respect to the sign of $\vec k_\nu \vec k_e=\pm 1$ there can be a
significant decrease (see also ref. {\cite{NSSV97}}) and increase of matter term.

Let us compare eqs. (\ref{V^Av=1}) and (\ref{V^Av=-1}) with eq. (\ref{V^Av1})
for the potential $V^A$  which is also valid for the transversal
polarization, $\vec \zeta_e \vec v_e=0$, and relativistic
motion of matter, $v_e \sim 1$. It follows that in the relativistic case and
transversal polarization of matter the potential $V^A$ is zero not for any
momenta of neutrinos (as it was claimed in \cite{NSSV97}), but only when
neutrino is also propagating in the transverse plane with respect to the matter
polarization and in addition to
$\vec \zeta_e \vec v_e=0$ the condition $\vec \zeta_e \vec
k_\nu=0$  is valid.

The total potential $A=V^V+V^A$ for arbitrary motion of matter in the direction
of the neutrino propagation, $\vec k_e \vec k_\nu=1$, as it follows from
eqs. (\ref{V^Vz0}), (\ref{fac1}) and (\ref{V^Av=1}), is
\begin{equation}
A=\sqrt2G_Fn_0\sqrt{{1-v_e} \over {1+v_e}}(1+\vec \zeta_e \vec k_\nu).
\end{equation}
The total potential $A$ in the relativistic case has a general
suppression factor $(1-v_e)^{1/2} \ll 1$, and is zero for the
complete matter polarization $\vec \zeta_e \vec k_{e}=-1$. In
the case of opposite motion of matter and the neutrino, $\vec k_e
\vec k_{\nu}=-1$, from eqs. (\ref{V^Vz0}), (\ref{fac2}) and
(\ref{V^Av=-1}) we get for the total potential,
\begin{equation}
A=\sqrt2G_Fn_0\sqrt{{1+v_e} \over {1-v_e}}(1+\vec \zeta_e \vec k_\nu).
\end{equation}
It follows that the total potential can be substantially increased
for the relativistic motion of matter due to the factor
$(1-v_e)^{-1/2} \gg 1$, however in the case of the longitudinal
polarization of matter, $\vec \zeta_e \vec k_{e}=1$, the
potential is zero.

In conclusion it is worth noting that a similar analysis can
be performed for any type of the neutrino flavour conversion and
different matter composition. As it follows from the above
considerations, there could be a decrease (increase) of the
particular matter component contribution to the difference of
neutrino potentials if the matter component is moving along
(against) the direction of neutrino propagation. If different
matter components move with different speeds, the difference of
neutrino effective potentials in the general case depends on each of
the speeds. As an example, let us consider the case of electron
neutrino $\nu_e$ conversion into sterile neutrino $\nu_s$ in the
presence of matter composed of electrons, protons and neutrons. We
also suppose that each of the matter components is characterised
by its total speed $\vec v_{e,p,n}$ and invariant number density
$n_0^{e,p,n}$. For simplicity we consider the unpolarized matter
components, then
\begin{equation}
\begin{array}{c}
\displaystyle A_{\nu_e \nu_s}={\sqrt2G_F \over 2}\left\{ {n^e_0(1+4\sin^2\theta_W)
\over \sqrt{1-v^2_e}}(1-\vec \beta \vec v_e) \right.+ \\ \left.
\displaystyle
{n^p_0(1-4\sin^2\theta_W)
\over \sqrt{1-v^2_p}}(1-\vec \beta \vec v_p)- {n^n_0 \over \sqrt{1-v^2_n}}(1-\vec \beta \vec v_n) \right\}.
\end{array}
\end{equation}
From this expression it follows that new phenomena are
expected to appear in the case of moving matter which are absent
if matter is at rest. For instance, it is easy to see that if $
\vec{v}_{e}\neq\vec{v}_{p},$ then the neutral current
contributions from the electron neutrino elastic forward
scattering off the background electrons and protons does not
cancel even in the case when the electron and proton invariant
densities are equal $n^{(0)}_e=n^{(0)}_p$. This phenomenon can exist in 
the relativistically expanding fireballs which are predicted within 
several models of gamma-ray bursts \cite {Mes02} and in which the electron 
Lorentz factor may exceed that of the proton by a factor up to the 
ratio of the proton to the electron mass.

\end{document}